\documentclass{ws-p10x7}
\begin{document}
\twocolumn[\hsize\textwidth\columnwidth\hsize\csname
@twocolumnfalse\endcsname
\vskip -0.5cm
\hfill \vbox{ \hbox{hep-ph/0010136} }\\
\vskip -0.5cm
\title{Solar and Atmospheric Neutrino Oscillations}
\author{M. C. Gonzalez-Garcia}
\address{Inst. de F\'{\i}sica Corpuscular \\
C.S.I.C. - Univ. de Val\`encia, Spain}
\maketitle
\abstract{\vskip -0.7cm
I review the status of neutrino masses and mixings
in the light of the solar and atmospheric neutrino data
in the framework of two--, three-- and four neutrino mixing.}]
\vskip -0.5cm
\section{Indications for Neutrino Mass: Two--Neutrino Analysis}
I \footnote{Talk given at ICHEP XXX, Osaka, July 2000.} 
first review the present experimental 
status for solar and atmospheric neutrinos and the results of the
different analysis in the framework of two--neutrino oscillations. 
\subsection{Solar Neutrinos}
\label{solar}
\begin{figure}[htbp]
\vskip -0.2cm
\begin{center}
\mbox{\epsfig{file=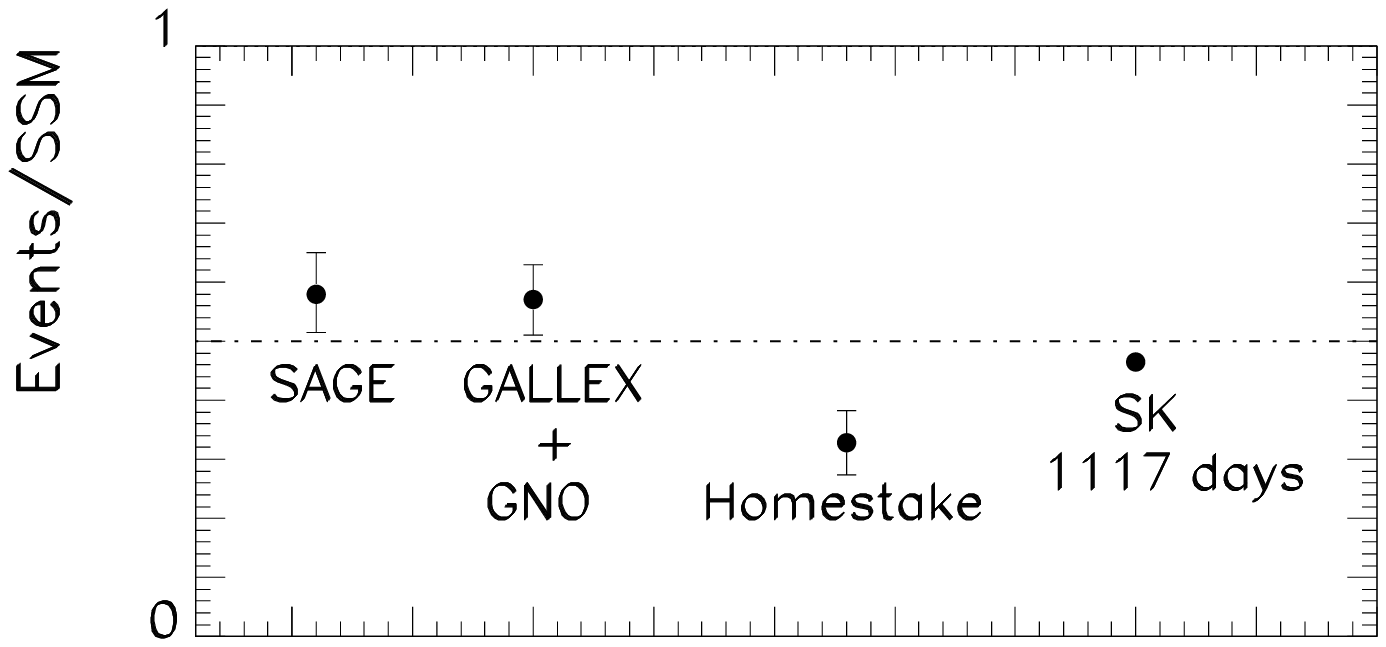,width=0.45\textwidth}}
\end{center}
\vskip -0.2cm
\label{rates}
\end{figure}
The sun is a source of $\nu_e's$ which are produced in the different
nuclear reactions taking place in its interior. Along this talk I will 
use the $\nu_e$ fluxes from Bahcall--Pinsonneault 
calculations~\cite{bp98}.
These neutrinos have been detected at the Earth by seven 
experiments which use different detection techniques \cite{sunexp00,sksol00}:
The chlorine experiment at Homestake, the water cerencov experiments 
Kamiokande and Super--Kamiokande (SK)  
and the radiochemical Gallex, GNO  and Sage experiments.
Due to the different energy threshold for the detection
reactions, these experiments  are sensitive to different parts of the 
solar neutrino spectrum. They all observe a deficit between 30 and 60 \% 
which seems to be energy dependent mainly due to the lower 
Chlorine rate. 
To the measurements of these six experiments we have to add also the
new results from SNO.
They are however still not in the form of definite measured rates which
could be included in this analysis.
  
SK has also presented their results after 1117 days of data taking on 
\cite{sksol00}:\\ 
-- The recoil electron energy spectrum: SK has measured the dependence
of the even rates on the recoil electron energy spectrum divided in
18 bins starting at 5.5 MeV. They have also reported the results
of a lower energy bin 5 MeV $<E_e<$5.5 MeV, but its systematic errors
are still under study and it is not included in their nor our analysis. 
The spectrum  shows no clear distortion with $\chi^2_{flat}=13/(17dof)$.\\
--The Zenith Angle Distribution (Day/Night Effect) which measures
the effect of the Earth Matter in the neutrino propagation.
SK finds few more events at night than 
during the day but the corresponding Day--Night asymmetry
$A_{D/N}=-0.034\pm 0.022 \pm 0.013$
is only 1.3$\sigma$ away from zero.

In order to combine both the Day--Night information and the
spectral data SK has also presented separately the measured recoil
energy spectrum during the day and during the night. This will be
referred in the following as the day--night spectra data which
contains $2\times 18$ data bins. 

\begin{figure}
\centerline{\protect\hbox{\psfig{file=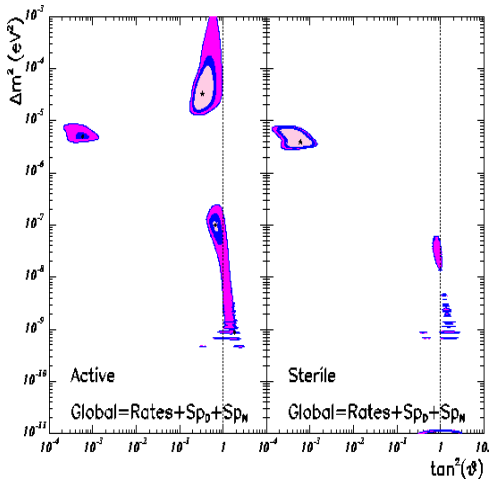,width=0.45\textwidth,height=0.3\textheight}}}
\vglue -0.1cm
\caption{Presently allowed solar neutrino parameters for two-neutrino
oscillations by the global analysis from Ref.~\protect\cite{nu2000}. 
The plotted regions are 90\%, 95  and 99\% CL.}
\label{solar2}
\end{figure}

The most generic and popular explanation of the solar neutrino anomaly is
in terms of neutrino masses and mixing leading to oscillations 
of $\nu_e$ into an active ($\nu_\mu$ and/or  $\nu_\tau$) or sterile 
neutrino, $\nu_s$. 
In Fig.~\ref{solar2} I show the allowed two--neutrino oscillation 
regions obtained in our updated global analysis of the solar neutrino data 
\cite{nu2000}. We show the possible 
solutions in the full parameter space for oscillations including both
MSW and vacuum, as well as quasi-vacuum oscillations (QVO) and matter effects 
for mixing angles in the second octant (the so called dark side).
These results have been obtained using
the general expression for the survival probability 
found by numerically solving the evolution equation 
in the Sun and the Earth matter valid in the full oscillation plane. 
In the case of active--active neutrino oscillations we find three 
allowed regions for the global fit: the SMA solution, the LMA and 
LOW-QVO solution. For sterile neutrinos only the SMA solution is 
allowed. For oscillations into an sterile neutrino there are 
differences partly due to
the fact that now the survival probability depends both on the electron and
neutron density in the Sun but mainly due to the lack of neutral current
contribution to the water cerencov experiments. 
In Table ~\ref{minima} I give
the values of the parameters in these minima as well as the GOF corresponding
to each solution.
\begin{table*}
\caption{Best fit points and GOF for the allowed solutions for  
combinations of observables.}
\label{minima}
\begin{center}
\begin{tabular}{|c|c|c|c|c|c|c|}
\hline
&  & \multicolumn{4}{c|}{Active}  & Sterile
\\\hline            
            & & SMA & LMA & LOW & VAC-QVO & SMA  \\\cline{2-7} 
   Rates    & $\Delta m^2$/eV$^2$ & $5.5\times 10^{-6} $  
            & $1.9\times 10^{-5} $  
            & $9.\times 10^{-8} $  
            & $9.7\times 10^{-11} $ 
            & $4.1\times 10^{-6} $     
                  \\ 
            & $\tan^2\theta$ & 0.0015     
            & 0.29   & 0.65    & 0.51 (1.94) & 0.0015  \\ \cline{2-7}             
            & Prob (\%)&  {50} \%  &  {8} \%  
            &  0.5 \% &  {2} \% &19 \%  \\\hline      
   Rates   &  $\Delta m^2$/eV$^2$ & $5.0\times 10^{-6} $  
            & $3.2\times 10^{-5} $  
            & $1.\times 10^{-7} $  
            & $8.6\times 10^{-10} $
            & $3.9\times 10^{-6} $   \\ 
+Spec$_D$   & $\tan^2\theta$ & 0.00058     
            & 0.33   & 0.67    & 1.5 (QVO) &0.0006 \\ \cline{2-7}             
+Spec$_N$   & Prob (\%)&  {34} \%  &  {59} \%  & {40} \% &  29 \%  &30\%  
\\\hline      
\end{tabular}
\end{center}
\end{table*}
There are some points concerning these results that
I would like to stress:\\
(a) Despite giving a worse fit to the observed total rates, once the 
day--night spectra data is included the LMA gives the best 
fit. This is mainly driven by the flatness of the spectrum and it was
already the case with the last year data.\\
(b) The GOF of the LOW solution has increased considerably as it describes
the spectrum data very well despite it gives a very bad fit to the
global rates. Notice also
that LOW and QVO regions are connected at the 99 \%CL and they 
extend into the second octant so maximal mixing is allowed at 99 \% CL 
for $\Delta m^2$ in the LOW-QVO region. \\
(c) As for the SMA the result from the correct statistically combined 
analysis shown in Fig.~\ref{solar2} and in Table ~\ref{minima} 
indicates  that the SMA can describe
the full data set with a probability of 34\% but it is now shifted
to smaller mixing angles to account for the flatter spectrum. \\
(d) Similar statement holds for the SMA solution for sterile neutrinos.\\
Thus the conclusion is that from the statistical point of view all 
solutions are acceptable since they all provide a reasonable GOF to the
full data set. 
LMA and LOW-QVO solutions for oscillations into active neutrino seem slightly 
favoured over SMA solutions for oscillations into active or sterile neutrinos 
but these last two are not ruled out. 
\subsection{Atmospheric Neutrinos}
Atmospheric showers are initiated when primary cosmic rays hit the
Earth's atmosphere. Secondary mesons produced in this collision,
mostly pions and kaons, decay and give rise to electron and muon
neutrino and anti-neutrinos fluxes.  
Atmospheric neutrinos can be detected in underground detectors by direct 
observation of their charged current interaction inside the detector.
These are the so called contained events. 
SK has divided their contained data sample into
sub-GeV events with visible energy below 1.2 GeV and multi-GeV above such
cutoff. On average, sub-GeV events arise from neutrinos of several hundreds of
MeV while multi-GeV events are originated by neutrinos with energies of the
order of several GeV. Higher energy muon neutrinos
and antineutrinos can also be detected indirectly by observing the muons
produced in their charged current interactions in the vicinity of the
detector. These are the so called upgoing muons. Should the muon 
stop inside the detector, it will be classified as a ``stopping'' muon,
(which arises from neutrinos of energies around ten GeV)
while if the muon track crosses the full detector the event is 
classified as a ``through-going'' muon which is originated by neutrinos
with energies of the order of hundred GeV.

At present the atmospheric neutrino anomaly (ANA) can be summarized in 
three observations: \\
-- There has been a long-standing deficit of about 60 \%  between the 
predicted and observed 
$\nu_\mu$$/\nu_e$ ratio of the contained events \cite{atmexp} now strengthened 
by the high statistics sample collected at the SK experiment \cite{skatm00}. 
\\
-- The most important feature of the atmospheric neutrino
data at SK is that it exhibits a {\sl zenith-angle-dependent} deficit of 
muon neutrinos which indicates that the deficit is larger for muon neutrinos
coming from below the horizon which have traveled longer distances 
before reaching the detector. \\
-- The deficit for thrugoing muons is smaller that
for stopping muons, {\it i.e.} the deficit decreases as the neutrino 
energy grows.

The most likely solution of the ANA involves neutrino
oscillations. In principle we can invoke various neutrino oscillation
channels, involving the conversion of $\nu_\mu$ into either $\nu_e$ or 
$\nu_\tau$
(active-active transitions) or the oscillation of $\nu_\mu$ into a sterile
neutrino $\nu_s$ (active-sterile transitions) \cite{2atmour}. 
Oscillations into electron neutrinos are nowadays ruled out since
they cannot describe the measured angular dependence of muon-like
contained events \cite{2atmour}. Moreover the most favoured range
of masses and mixings for this channel have been excluded by the 
negative results from the CHOOZ reactor experiment \cite{chooz}.

\begin{figure}
\centerline
{\protect\hbox{\psfig{file=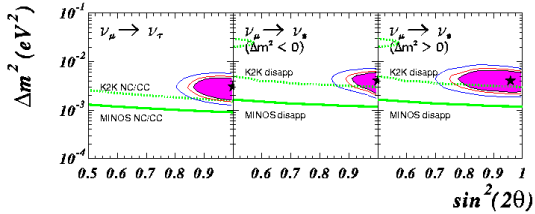,width=0.5\textwidth}}}
\caption{Allowed atmospheric oscillation parameters all for 
experiments, combined at
90 (shadowed area), 95 \% and 99 \% CL (thin solid line) for all possible
oscillation channels, from Ref.~\protect\cite{2atmour}.  
The expected sensitivity for upcoming long-baseline experiments is also 
displayed.}
\label{atmos2} 
\end{figure}

In Fig.~\ref{atmos2} I show the allowed neutrino oscillation parameters
obtained in our global fit \cite{2atmour}  of the full data set of atmospheric
neutrino data on vertex contained events at IMB, Nusex, Frejus,
Soudan, Kamiokande \cite{atmexp} and SK experiments \cite{skatm00} 
as well as upward going muon data from SK, Macro and Baksan experiments
in the different oscillation channels.

The two panels corresponding to oscillations into sterile neutrinos  
in Fig.~\ref{atmos2} differ in the sign of the $\Delta m^2$ which was 
assumed in the analysis of the matter effects in the
Earth for the $\nu_\mu \to \nu_s$ oscillations. 
Concerning the quality of the fits our results show 
that the best fit to the full sample is obtained for the 
$\nu_\mu \to \nu_\tau$ channel although from the global analysis
oscillations into sterile neutrinos cannot be ruled out at any reasonable CL. 
Due to matter effects the distribution for upgoing muons in the case of 
$\nu_\mu \to \nu_s$ are flatter than for $\nu_\mu \to \nu_\tau$. 
Data show a somehow steeper angular dependence which can be better
described by $\nu_\mu \to \nu_\tau$. This leads to the better quality
of the global fit in this channel. Pushing further this feature 
SK  collaboration has presented an 
analysis of the angular  dependence of the through-going muon data in 
combination with the up-down asymmetry of partially contained events 
and the neutral current enriched events
which seems to disfavour the possibility $\nu_\mu \to \nu_s$ at 
the 3--$\sigma$ level \cite{skatm00}.
\section{Three--Neutrino Oscillations}
In the previous section I have discussed the evidences for neutrino
masses and mixings as usually formulated in the two--neutrino oscillation
scenario. Let us now fit all the different evidences in a 
common three--neutrino framework and see what is our present knowledge 
of the neutrino mixing and masses. Here I present a brief summary of 
such analysis performed in Ref.~\cite{ourthree} and I refer to that 
publication for further details and references.

The evolution equation for the three neutrino flavours can be written as:
\begin{equation}
-i\frac{d\nu}{dt}=\left[U \frac{M_\nu}{2 E} U^\dagger +H_{int}\right ] \; ,
\label{evolution}
\end{equation}
where $M_\nu$ is the diagonal mass matrix for the three neutrinos and
$U$ is the unitary matrix relating the flavour and the mass basis.
$H_{int}$ is the Hamiltonian describing the neutrino interactions.
In general $U$ contains 3 mixing angles and 1 or 3 CP violating phases
depending on whether the neutrinos are Dirac or Majorana. I will neglect 
the CP violating phases as they are not accessible by the existing
experiments.  
In this case the mixing matrix can be conveniently chosen in the form
\begin{equation}
U=R_{23}(\theta_{23})\times R_{13}(\theta_{13}) \times R_{12}(\theta_{12})\; ,
\end{equation}
where $R_{ij}$ is a rotation matrix in the plane $ij$.
With this the parameter set relevant for the joint study of solar and
atmospheric conversions becomes five-dimensional:
\begin{equation} \begin{array}{cc} 
   \label{oscpardef}
    \Delta m^2_{\odot} \equiv \Delta m^2_{21}\; ,\;& 
    \Delta m^2_{atm}  \equiv \Delta m^2_{32}, \\
    \theta_{\odot}     \equiv \theta_{12} \; , \;&
    \theta_{atm}        \equiv \theta_{23}\; , \; 
    \theta_{reac}   \equiv \theta_{13}\; ,
\end{array} 
\end{equation}
where all mixing angles are assumed to lie in the full range from
$[0,\pi/2]$. 

In general the transition probabilities will present an oscillatory
behaviour with two oscillation lengths. However 
from the required hierarchy in the splittings $\Delta m^2_{atm} \gg
\Delta m^2_{\odot}$ indicated by the solutions to the solar and
atmospheric neutrino anomalies it follows that:\\
-- For solar neutrinos the oscillations with the atmospheric oscillation 
length are averaged out and the survival probability takes the form:
\begin{equation}
P^{3\nu}_{ee,MSW}
=\sin^4\theta_{13}+ \cos^4\theta_{13}P^{2\nu}_{ee,MSW} 
\label{p3}
\end{equation}
where $P^{2\nu}_{ee,MSW}$ is obtained with the modified sun density
$N_{e}\rightarrow \cos^2\theta_{13} N_e $. 
So the analyses of solar data constrain
three of the five independent oscillation parameters: $\Delta
m^2_{21}, \theta_{12}$ and $\theta_{13}$. \\
-- Conversely for atmospheric neutrinos, the solar wavelength is too long
and the corresponding oscillating phase is negligible. As a consequence
the atmospheric
data analysis restricts $\Delta m^2_{32}$, $\theta_{23}$ and
$\theta_{13}$, the latter being the only parameter common to both solar 
and atmospheric neutrino oscillations and
which may potentially allow for some mutual influence.

Therefore solar and atmospheric neutrino oscillations
decouple in the limit $\theta_{13}=0$. In this case the values of 
allowed parameters can be obtained directly
from the results of the analysis in terms of two--neutrino oscillations 
presented in the first section.
Deviations from the two--neutrino scenario are then determined by the
size of the mixing $\theta_{13}$. This angle is constrained by
the CHOOZ reactor experiment which imposes an strong lower
limit on the probability
\begin{equation}
P_{ee}^{CHOOZ}=1 -\sin^2(2\theta_{13})
\sin^2( \frac{\Delta m^2_{32} L}{4 E_\nu})
\label{pchooz}
\end{equation}
$> 0.91$ at 90 \% CL for $\Delta m^2_{32}>10^{-3}$ eV$^2$.
\begin{figure}
\centerline{\protect\hbox{\psfig{file=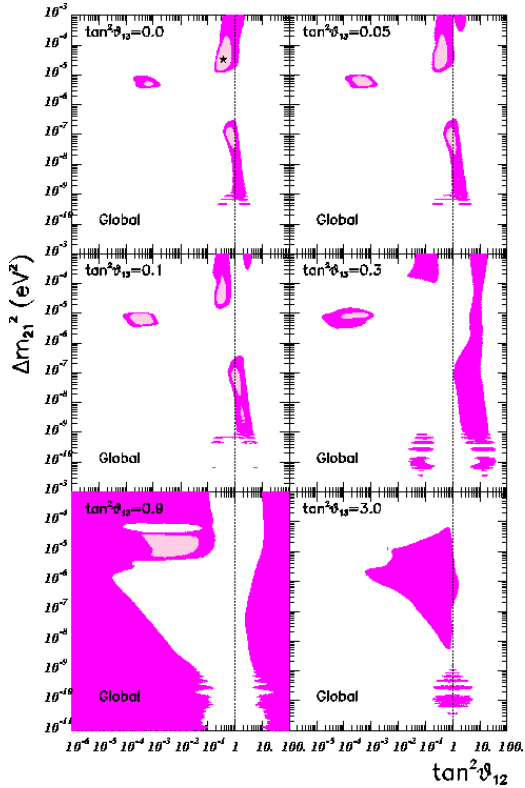,width=0.48\textwidth}}}
\caption{Allowed regions at 90\% and 99\% CL for the oscillation parameters 
$\Delta m_{21}$ and $\tan^2(\theta_{12})$ 
from the global analysis of the solar neutrino  data
in the framework of three--neutrino oscillations
for different values of the angle $\theta_{13}$ 
from Ref.~\protect\cite{ourthree}.}
\label{solar3} 
\end{figure}
 
The first question to answer is how
the presence of this new angle affects the analysis of the solar
and atmospheric neutrino data \cite{ourthree}. 
In Fig.~\ref{solar3} I show the allowed regions for the oscillation 
parameters $\Delta m_{21}$
and $\tan^2(\theta_{12})$ from our global analysis of the solar neutrino 
data  in the framework of three--neutrino oscillations
for different values of the angle $\theta_{13}$.
The allowed regions  
for a given CL are defined as the set of points satisfying  
the condition
$\chi^2(\Delta m_{12}^2,\tan^2\theta_{12},\tan^2\theta_{13})
-\chi^2_{min}\leq \Delta\chi^2 \mbox{(CL, 3~dof)} $
where, for instance, $\Delta\chi^2($CL, 3~dof)=6.25, 7.83, and 11.36 for 
CL=90, 95, and 99 \% respectively. The global minimum 
used in the construction of the regions lays in the LMA region and
corresponds to $\tan^2\theta_{13}=0$, this is, for the ``decoupled'' 
scenario. Notice that the only difference between the first panel 
in Fig.~\ref{solar3} and the  active oscillations solution in 
Fig.~\ref{solar3} is due to the different numbers of dof used 
in the definition of the regions.
The behaviour of the regions illustrate the ``tension'' between 
the data on the total event rates
which favour smaller $\theta_{13}$ values and the day--night spectra
which allow larger values. It can also be understood
as the ``tension'' between the energy dependent and constant pieces of
the electron survival probability in Eq.~(\ref{p3}).

As seen in the figure the effect is small unless large values
of $\theta_{13}$ are involved. From Fig.~\ref{solar3} we find that as 
$\tan^2\theta_{13}$
increases all the allowed regions disappear, leading to an upper bound 
on $\tan^2\theta_{13}$ for any value of $\Delta m^2_{21}$, independently 
of the values taken by
the other parameters in the three--neutrino mixing matrix. 
The corresponding 90 and 99 \% CL bounds are tabulated in
Table~\ref{table:limits}.
\begin{table}
    \begin{tabular}{c|cc}
Data Set & \multicolumn{2}{c}{$\tan^2\theta_{13}$}  \\       
         & min   & limit 99\% \\
        \hline
        Solar        & 0.0   & 3.5 ($62^\circ$)   \\
        Atmos        & 0.026 & 0.57 ($37^\circ$)   \\
        Atm+CHOOZ   & 0.005   & 0.08 ($16^\circ$) \\
        Atm+Solar   & 0.015   & 0.52 ($36^\circ$)     \\
        Atm+Solar+Chooz  & 0.005 & 0.085 ($16^\circ$) 
    \end{tabular}
\label{table:limits}
\end{table}

\begin{figure}
\begin{center} 
\mbox{\epsfig{file=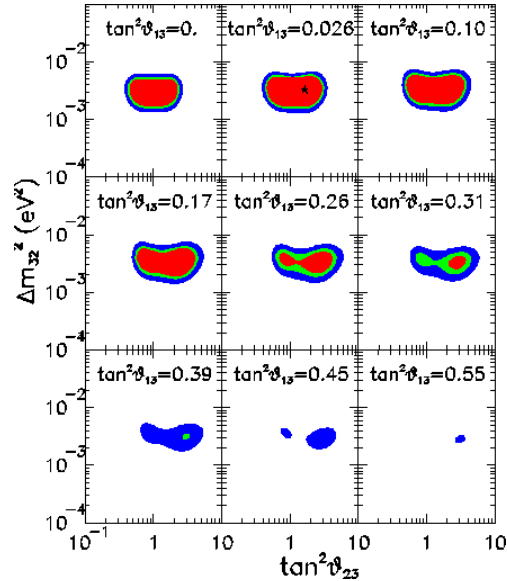,width=0.46\textwidth}} 
\end{center} 
    \vspace{3mm}
    \caption{90, 95 and 99\% CL
      three--neutrino allowed regions in $(\tan^2\theta_{23}, ~\Delta
      m^2_{32})$ for different $\tan^2\theta_{13}$ values, for the
      combination of global analysis of atmospheric neutrino data
      from Ref.~\protect\cite{ourthree}.  The best--fit point is
      denoted as a star.}
    \label{atmos3}
\end{figure}

As for the atmospheric neutrino data in
Fig.~\ref{atmos3} I show the $(\tan^2\theta_{23}, ~\Delta
m^2_{32})$ allowed regions, for different values of
$\tan^2\theta_{13}$ from the global analysis of the atmospheric
neutrino data. 
The upper-left panel, $\tan^2\theta_{13} = 0$, corresponds to
pure $\nu_\mu \to \nu_\tau$ oscillations, and one can note the exact
symmetry of the contour regions under the transformation $\theta_{23}
\to \pi/4 - \theta_{23}$. This symmetry follows from the fact that in
the pure $\nu_\mu \to \nu_\tau$ channel matter effects cancel out and
the oscillation probability depends on $\theta_{23}$ only through the
double--valued function $\sin^2(2\theta_{23})$.  For non-vanishing
values of $\theta_{13}$ this symmetry breaks due to the
three--neutrino mixing structure even if matter effects are neglected
We see that the analysis of the full atmospheric neutrino 
data in the framework of three--neutrino oscillations clearly favours
the $\nu_\mu \to \nu_\tau$ oscillation hypothesis. As a matter of fact 
the best fit corresponds to a small value of $\theta_{13}= 9^\circ$.
With our sign assignment we find that for non-zero values
of $\theta_{13}$ the allowed regions become larger in the second
octant of $\theta_{23}$. No region of parameter space is 
allowed (even at 99\% C.L.)  for $\tan^2\theta_{13} > 0.6$.  
Larger values of $\tan^2\theta_{13}$ would imply a too large contribution
of $\nu_\mu \to \nu_e$ and would spoil the description of the angular
distribution of contained events.
The mass difference relevant for the atmospheric
analysis is restricted to lay in the interval: 
$1.25\times 10^{-3}<\; \Delta m^2_{32}/\mbox{\rm eV$^2$}<8\times 10^{-3}$
at 99 \% CL. Thus it is within the range of sensitivity of the CHOOZ experiment
and as a consequence the angle $\theta_{13}$ is further constrained when
we include in the analysis the results from this reactor experiment.
This is illustrated in Table\ref{table:limits} 
where one sees that the limit on $\tan^2\theta_{13}$
is strengthen when the CHOOZ data is combined with the atmospheric 
neutrino results. 

One can finally perform a global analysis in the 
five dimensional parameter space combining the full set of  solar, 
atmospheric and reactor data. As an illustration of such analysis I 
present in Table\ref{table:limits} the resulting bounds on $\theta_{13}$.
The final results from the joint solar, atmospheric, and
reactor neutrino data analysis lead to the following allowed ranges of
parameters at 99\% CL
\begin{eqnarray}
1.1\times 10^{-3}<& \Delta m^2_{32}/\mbox{\rm eV$^2$}&< 
7.3\times 10^{-3} \\ \nonumber
0.33<&\tan^2\theta_{23}&< 3.8 \\\nonumber
\tan^2\theta_{13} &< 0.085 &\;\; \mbox{(if solar LMA)} 
\\\nonumber 
\tan^2\theta_{13} &< 0.135 &\;\;  \mbox{(if solar SMA)} \nonumber  
\label{globalranges}
\end{eqnarray} 
In conclusion we see that from our statistical analysis of the solar
data it emerges that the status of the large mixing--type solutions
has been further improved with respect to the previous  SK
data sample, due mainly to the substantially flatter recoil electron
energy spectrum. In contrast, there has been no fundamental change,
other than further improvement due to statistics, on the status of the
atmospheric data.  For the latter the oscillation picture clearly
favours large mixing, while for the solar case the preference is still
not overwhelming. Both solar and atmospheric data favour small
values of the additional $\theta_{13}$ mixing and this behaviour is
strengthened by the inclusion of the reactor limit. 
\section{Four--Neutrino Oscillations}
Together with the results from the solar and atmospheric neutrino  
experiments we have one more evidences pointing out towards the existence of  
neutrino masses and mixing: the LSND 
results. These three evidences can be accommodated 
in a single neutrino oscillation framework only if there are at least 
three different scales of neutrino mass-squared differences which
requires the existence of a light sterile neutrino.
Here I present a brief update of the analysis performed 
in Ref.~\cite{four} of solar neutrino data in such framework of
four--neutrino mixing and I refer to that publication for further 
as well as for the relevant references. 

In four-neutrino schemes the rotation $U$ relating the flavor neutrino fields 
to the mass eigenstates fields is a $4{\times}4$ unitary mixing matrix, 
which contains, in general
6 mixing angles (I neglect here the CP phases). 
Existing bounds from negative searches for neutrino oscillations performed
at collider as well as reactor experiments impose severe constrains
on the possible mass hierarchies as well as mixing structures for the
four--neutrino scenario. In particular they imply:\\
(a) Four-neutrino schemes with 
two pairs of close masses separated by a gap of about 1 eV 
which gives the mass-squared difference responsible for the 
oscillations observed in the LSND experiment, 
can accommodate better the results of all neutrino oscillation experiments. \\
(b) In the study of solar and atmospheric neutrino oscillations 
only four mixing angles are relevant and the $U$ 
matrix can be written as
$U = R_{34} \, R_{24} \, R_{23} \, R_{12} $
We choose  solar neutrino oscillations 
to be  generated by the mass-square difference 
between $\nu_2$ and $\nu_1$. With this choice the survival 
of solar $\nu_e$'s mainly depends on the mixing angle 
$\vartheta_{12}$ and it is independent of $\vartheta_{34}$.  
The mixing 
$\vartheta_{23}$ and $\vartheta_{24}$ 
determine the relative amount of transitions into sterile $\nu_s$ 
or active $\nu_\mu$ and $\nu_\tau$ only through the combination 
$\cos{\vartheta_{23}} \cos{\vartheta_{24}}$ $(c^2_{23} c^2_{24})$. 
We distinguish the following limiting cases: \\
$\bullet$ $c^2_{23} c^2_{24} = 0$ corresponding to the limit of 
pure two-generation 
$\nu_e\to\nu_a$ transitions.\\
$\bullet$ $c^2_{23} c^2_{24}= 1$ 
for which we have the limit of 
pure two-generation 
$\nu_e\to\nu_s$ transitions.\\
$\bullet$ If $c^2_{23} c^2_{24}\neq 1$, 
solar $\nu_e$'s can transform 
in the linear combination $\nu_a$ of active $\nu_\mu$ and $\nu_\tau$. 

In the general case of simultaneous $\nu_e\to\nu_s$ and 
$\nu_e\to\nu_a$ oscillations 
the corresponding probabilities are given by 
\begin{eqnarray} 
&& 
P_{\nu_e\to\nu_s} 
= 
c^2_{23} c^2_{24} 
\left( 1 - P_{\nu_e\to\nu_e} \right) 
\,, 
\label{Pes} 
\label{Pea} 
\\ 
&& 
P_{\nu_e\to\nu_a} 
= 
\left( 1 - c^2_{23} c^2_{24} \right) 
\left( 1 - P_{\nu_e\to\nu_e} \right) 
\,. 
\end{eqnarray} 
where  $P_{\nu_e\to\nu_e}$ takes the standard two--neutrino oscillation
for $\Delta m^2_{21}$ and $\theta_{12}$ but computed for the modified 
matter potential  $A \equiv A_{CC} + c^2_{23}c^2_{24} A_{NC}$ 
Thus the analysis of the solar neutrino data in the
four--neutrino mixing schemes is equivalent to the two--neutrino
analysis but taking into account that the parameter space is now 
three--dimensional $(\Delta m^2_{12},\tan^2\vartheta_{12}, 
c^2_{23}c^2_{24})$. 

\begin{figure}[htbp]
\vskip -0.2cm
\begin{center}
\mbox{\epsfig{file=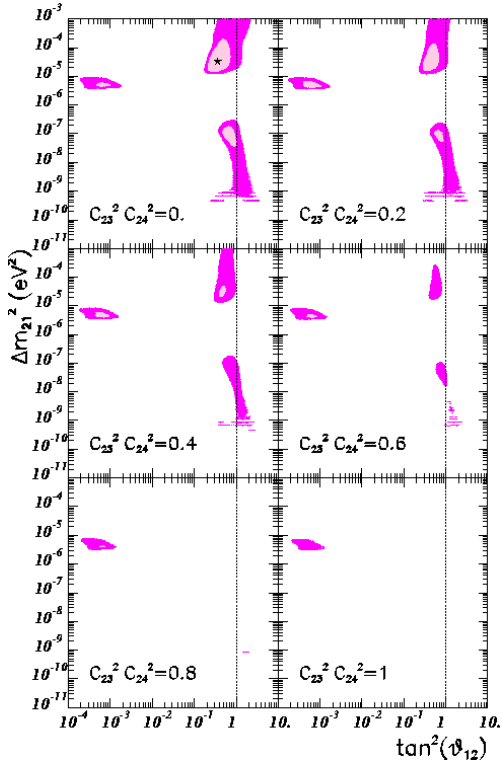,width=0.46\textwidth}}
\end{center}
\vskip -0.2cm
\caption{Results of the global analysis for the allowed regions in   
$\Delta{m}^2_{21}$ and $\sin^2 \vartheta_{12}$  
for the four--neutrino oscillations. 
The different panels 
represent the allowed regions at 99\% (darker) and 90\% CL (lighter).  
The best--fit point in the three parameter space is  
plotted as a star.} 
\label{four}
\end{figure}

I first present the results of the allowed regions in the three--parameter 
space for the global combination of observables. 
In  Fig.~\ref{four} I show the  
sections of the three--dimensional allowed  volume in the plane 
($\Delta{m}^2_{21},\tan^2(\vartheta_{12})$) for different values of 
$c_{23}^2c_{24}^2$.  The global minimum used in the construction of the 
regions lies in the LMA region and for pure active oscillations 
value of $c_{23}^2c_{24}^2=0$. 
As seen in  Fig.~\ref{four} 
the SMA region is always a valid solution  
for any value of $c_{23}^2c_{24}^2$. This is expected as  
in the two--neutrino oscillation picture this solution holds both  
for pure active--active and pure active--sterile oscillations.
Notice, however, that the statistical analysis is different: 
in the two--neutrino picture the pure active--active and active--sterile 
cases are analyzed separately, 
whereas in the four--neutrino picture they are taken into account 
simultaneously in a consistent scheme. We see that in this ``unified'' 
framework, since the 
GOF of the SMA solution for pure sterile oscillations is worse than
for SMA pure active oscillations (as discussed in the first section), 
the corresponding allowed region is smaller as they are now defined 
with respect to a common minimum.

On the other hand, the LMA, LOW and QVO solutions disappear for 
increasing values of the mixing $c_{23}^2c_{24}^2$. 
I list in Table \ref{limits4} the maximum allowed values of
$c_{23}^2c_{24}^2$ for which each of the solutions is allowed at 
a given CL. We see that at 95 \% CL the LMA solution is allowed 
for maximal active-sterile mixing $c_{23}^2c_{24}^2=0.5$ while 
at 99\% CL all solutions are possible for this maximal mixing case.
\begin{table}
\caption{Maximum allowed value of $c^2_{23} c^2_{24}$ 
for the different solutions to the solar neutrino problem}.  
\label{limits4}
\begin{center}
\begin{tabular}{c|cccc}
CL  & SMA & LMA & LOW & QVO \\ \hline
90  & 0.9 & 0.44 & 0.3 & forbidden \\
95  & all & 0.53 & 0.44 & 0.28 \\
99  & all & 0.72 & 0.77 & 0.88 
\end{tabular}
\end{center}
\end{table}

This work was supported by grants DGICYT-PB98-0693 and PB97-1261, 
GV99-3-1-01, and ERBFMRXCT960090 and
HPRN-CT-2000-00148 of the EU.

\end{document}